\documentclass[aps,prl,twocolumn,showpacs]{revtex4}

\usepackage{graphicx}
\usepackage{times}

\begin{document}

\title{Comment on "Violation of Anderson's Theorem for the sign-reversing
s-wave state of Iron-Pnictide Superconductors \cite{Kontani}"}

\author{Yunkyu Bang}
\affiliation{Department of Physics,
Chonnam National University, Kwangju 500-757, Republic of Korea\\
and Asia Pacific Center for Theoretical Physics, Pohang 790-784,
Republic of Korea}


\maketitle

In Ref.\cite{Kontani}, Onari {\it et al.} studied the nonmagnetic
impurity effects on the sign-changing s-wave state ($s_{\pm}$) in
the Fe-based superconductors and claimed: (1) the orbital-less
model such as a band basis model has no impurity pair breaking
effect in the unitary limit; (2) however, the model with the
orbital degree of freedom can have the impurity pair breaking
effect as strong as in the nodal-gap superconductors, therefore
the presence of the orbital degree of freedom is essential to
describe the correct impurity effects. In this comment, we point
out that the claims (1) and (2) are incorrect conclusions and show
that both the band basis model and the orbital basis model have
the same pair breaking effect and the presence or absence of the
orbital degree of freedom is irrelevant for the impurity effects.

The authors of Ref.[1] showed that the $\mathcal{T}$-matrix in the
band basis $\hat{\mathcal{T}}^b$ always becomes band diagonal when
the impurity potential strength $I \rightarrow \infty$, hence the
pair breaking interband scattering process vanishes. The reasoning
for this is that the diagonal terms in $\hat{\mathcal{T}}^b$ are
always higher order in $I$ than the off-diagonal terms. The error
has occurred because Ref.[1] didn't subtract the bare impurity
potential $\hat{I}^b$ from $\hat{\mathcal{T}}^b$, which would
cause only the chemical potential shift and should be absorbed
into the redefined chemical potential. With this subtraction, the
diagonal and off-diagonal terms in $\hat{\mathcal{T}}^b$ always
become the same order in $I$, therefore there is a pair breaking
in the band basis $\mathcal{T}$-matrix for all values of $I$.

Then the authors of Ref.[1] introduced a five orbital model with
an impurity potential $\hat{I}^o = I \delta_{j,l}$ that is orbital
diagonal and momentum independent. As the authors showed, this
potential can be transformed into the band basis via an unitary
transformation as $U^{\dagger}\hat{I}^o U = \hat{I}^b$. Through
the unitary transformation, the momentum independence of
$\hat{I}^o$ continues to survive in $\hat{I}^b$ and it means that
the intraband and interband scattering terms in $\hat{I}^b$ are
equal strength \cite{note}. Therefore we can use the band basis
impurity formalism of Ref.[2] where the equal strength of the
intraband and interband impurity scattering was assumed for
simplicity, and every results of Ref.[1] can be reproduced,
clearly demonstrating that the presence or absence of the orbital
degree of freedom is irrelevant for the impurity effects contrary
to the main claim of Ref.[1].

\begin{figure}
\noindent
\includegraphics[width=80mm]{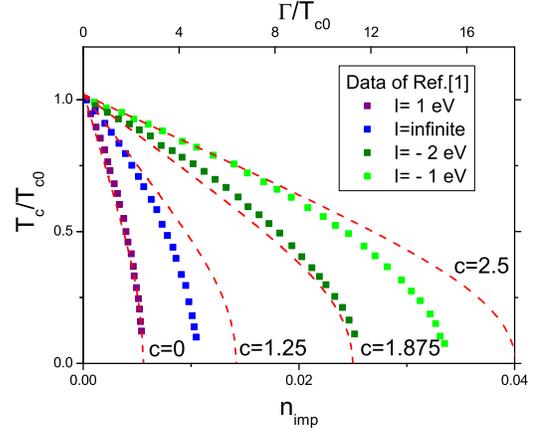}
\vspace{-0.6cm}
\caption{(Color online) Comparison of normalized $T_c$ data
for $s_{\pm}$-wave state from two model calculations.
Symbols are the data from Ref.[1] (bottom x-axis with $n_{imp}$)
and dashed lines are the calculations with the two band model\cite{Bang-imp}
(upper x-axis with $\Gamma/T_{c0}$) . \label{fig1}}
\end{figure}

The mapping of the particle-hole asymmetric five orbital model of
Ref.\cite{Kontani} to the particle-hole symmetric two band model
of Ref.\cite{Bang-imp} only needs to estimate the renormalized
impurity potential $I_{eff}$ due to the particle-hole asymmetry
and the total density of states $N_{tot}$. The particle-hole
asymmetry yields non-zero $g_3(\omega)$\cite{note2} which enters
in the combination of $(I^{-1} - g_3)$, hence it only renormalizes
the bare impurity potential $I$ to the effective one $I_{eff}$ as
$I_{eff}^{-1} = (I^{-1} - g_3)$ \cite{note3}. $g_3$ is weakly
frequency dependent and we can easily read off $g_3(\omega=0)
\approx 1 eV^{-1}$ from the fact that $I=1 eV$ produced the
maximum impurity scattering effect, i.e. $I_{eff} = \infty$ in
Ref.[1]. Now we obtain $I_{eff} (I=1,\infty,-2, -1 eV) = \infty,
1, -1.5$, and $-2 eV$, respectively, and understand the
non-monotonous relation between the pair breaking effect and the
values of $I$ in Ref.[1].

With $I_{eff} (I)$ and choosing $\pi N_{tot}=0.8/eV$ as a fitting
parameter, we calculated the $T_c$ suppression due to the
non-magnetic impurities for the $s_{\pm }$-wave state using the
two band model of Ref.[2], where the impurity potential strength
is parameterized by the phase shift parameter $c$ defined as
$\frac{1}{c} = \pi N_{tot}I_{eff}$. We obtained $|c|(I=1, \infty,
-2, -1 eV)= 0, 1.25, 1.875$, and $2.5$, respectively. Figure.1
shows the comparison between our calculations and the results of
the five orbital model of Ref.[1]. Two data sets closely track
each other. This result demonstrates the equivalence of the band
basis model and the orbital basis model and the irrelevance of the
presence or absence of the orbital degree of freedom for the
impurity effects in the Fe-based superconductors.

This work was supported by the grant NRF-2010-0009523 funded by
the National Research Foundation of Korea.


\begin{references}

\bibitem{Kontani}
S. Onari and H. Kontani, Phys. Rev. Lett. {\bf 103}, 177001
(2009).

\bibitem{Bang-imp}
Y. Bang, H.-Y. Choi, and H. Won,  Phys. Rev. B, {\bf 79}, 054529
(2009).

\bibitem{note}
This is indeed stated in Ref.[28] of the Ref.[1].

\bibitem{note2}
The $\sigma_3$ component of the local Nambu Green function $g_3
(\omega)= \frac{1}{N} \Sigma_k G_3 (k,\omega)$.

\bibitem{note3}
This implies that the particle-hole asymmetry can be always
absorbed into the renormalized impurity potential $I_{eff}$ and
can be ignored for all practical purpose.
\end{references}
\end{document}